\documentclass[apjl]{emulateapj}

\voffset= - 1 true cm

\shorttitle{Fundamental Galactic Parameters}

\shortauthors{Majewski, et al.}

\begin{document}


\title{Measuring Fundamental Galactic Parameters with Stellar Tidal Streams and SIM PlanetQuest}


\author{Steven R. Majewski\altaffilmark{1},
 David R. Law\altaffilmark{2}, Allyson A. Polak\altaffilmark{1} \& Richard J. Patterson\altaffilmark{1}} 
\altaffiltext{1}{Dept. of Astronomy, Univ. Virginia, Charlottesville, VA 22903}
\altaffiltext{2}{Caltech, Dept. of Astronomy, MS 105-24, Pasadena, CA 91125}
 
\email{srm4n, aap5u,rjp0i@virginia.edu, drlaw@astro.caltech.edu}

\begin{abstract}

Extended halo tidal streams from disrupting Milky Way satellites offer new
opportunities for gauging fundamental Galactic parameters without
challenging observations of the Galactic center.  In the roughly spherical 
Galactic potential tidal debris from a satellite system is largely confined to a 
single plane containing the Galactic center, so accurate distances to stars in the 
tidal stream can be used to gauge the Galactic center distance, $R_{0}$, given 
reasonable projection of the stream orbital pole on the $X_{GC}$ axis. 
Alternatively, a tidal stream with orbital pole near the $Y_{GC}$ axis, like 
the Sagittarius stream,
can be used to derive the speed of the Local Standard of Rest ($\Theta_{\rm LSR}$).
Modest improvements in current astrometric catalogues might allow this measurement to be made, but
NASA's Space Interferometry Mission (SIM PlanetQuest) can definitively obtain both $R_{0}$ and
$\Theta_{\rm LSR}$ using tidal streams.


\end{abstract}

\keywords{Milky Way: structure -- Milky Way: dynamics -- Sagittarius dwarf galaxy}

\section{Distance to the Galactic Center}

With the assumption that globular clusters trace the general shape and extent
of the Milky Way (MW), Shapley (1918) first showed how they can be used to estimate the distance ($R_{0}$) to the Galactic center (GC), expected to lie at the center of the cluster
distribution.  Though Shapley's first execution of this experiment exaggerated 
$R_{0}$ due to cluster distance scale problems, the overall scheme 
of mapping an extended distribution of Galactic tracer objects to determine the location of its 
center remains a valid, if traditionally underutilized, strategy.

The globular cluster sample is relatively small
and concentrated to the GC, where dust effects introduce large distance
uncertainties and a likely still incomplete and lop-sided cluster census.
Population II tracers like RR Lyrae, blue horizontal branch (BHB) or giant stars are much more
plentiful outside of the MW bulge.
Unfortunately, the current census for these tracers is even more incomplete than for globulars.  Though this
situation may be remedied by currently planned wide angle surveys,
several inherent problems remain with exploitation of these tracers as GC benchmarks.
As with the clusters, the MW Zone of Avoidance (ZA) will always
result in biased sample distributions
and potential $R_{0}$ underestimates  ---
exacerbated if surveys do not 
reach the far side of the MW.  Even more challenging 
is that the global distributions of halo stars are far from dynamically mixed: 
Recent surveys of the above tracers reveal a halo 
streaked with substructure, 
likely originating as satellite disruption debris 
(e.g., Vivas et al. 2001, Newberg et al. 2002, Majewski 2004) and eroding
simple halo axisymmetry. 


The very  existence of 
numerous tidal streams 
motivates the present contribution. {\it Individual} tidal streams actually possess a relatively simple
spatial configuration. Within spherical potentials, 
tidal debris arms from a disrupting satellite will lie along 
the satellite orbital plane, which contains the GC.  
A sufficiently extended 
tidal debris arc defines that
plane, which intersects the MW $X_{GC}$ axis at the GC.
This simpler, almost two-dimensional
geometry of tidal stream arcs removes 
the need for 
sample completeness: In principle, $R_{0}$  should be derivable from
the $(l,b,$distance) distribution of only a large enough sample of tidal stream
stars to define their orbital plane.\footnote{Samples
should be unbiased with respect to spread perpendicular to that plane, but
this should be trivial to achieve.}   

In reality, non-spherical
potentials precess tidal streams.  Fortunately, this is a relatively small effect in the MW, 
at least for $R_{GC}$'s of 
tens of kiloparsecs.  
Johnston et al. (2005; ``J05") showed the Sagittarius (Sgr) tidal stream precession is sufficiently
small to conclude the MW potential is only slightly oblate
within the Sgr orbit (peri:apo-Galacticon of 13:57 kpc).
Moreover, as pointed out also by Helmi (2004), that part of the Sgr trailing arm arcing
across the southern MW hemisphere (see Majewski
et al. 2003, ``MSWO'') is so
dynamically young that it hasn't had {\it time} to precess 
(see Fig.\ 5 of J05).  
Unfortunately, as noted by MSWO, 
Sgr is in almost the worst possible orientation to undertake
the proposed $R_{0}$-gauging: With a virtually negligible angle between the Sgr orbital
plane and $X_{GC}$ axis, small errors in the definition of the orbital plane (due to 
small residual precession, and the finite width of the debris plane) lead to 
substantial uncertainties in derivation of $R_{0}$.  

The ideal tidal debris configuration for estimating $R_{0}$ has a {\it pole} 
closer to the $X_{GC}$ axis. 
Given the pace of discovery, such a stream may soon be found.  
Based on the nearly polar orientation of the HI Magellanic Stream and the typically measured 
proper motions ($\mu$'s)
 for the Magellanic Clouds (Gardiner \& Noguchi 1996, van der Marel et al. 2002 and references therein),
it is clear that a stellar counterpart to the Magellanic Stream
would have almost the perfect orientation for gauging $R_{0}$.  

Systematic errors in a tracer distance scale translate to estimates of $R_{0}$. 
However, because streams contain different stellar types 
(e.g., giant stars, RR Lyrae, BHB), 
uncertainties from photometric/spectroscopic
parallaxes can be cross-checked.
In most cases, reddening and crowding effects can be of negligible concern.
Alternatively, with NASA's Space Interferometry Mission (SIM), direct {\it trigonometric}
parallaxes will be well within reach: For a putative Magellanic stellar stream orbiting 
at $\sim50$ kpc radius, the $\sim10$-20 $\mu$as parallaxes are well above the SIM 
wide-angle astrometric accuracy goal of 4$\mu$as, assuming K giant star tracers ($V \sim 18$). 

As a test of what might be achieved, we ran N-body simulations of different mass satellites
disrupting for 5 or 10 Gyr (whatever was needed to produce $>270^{\circ}$-long tails)
in the Galactic potential that best fitted the Sgr debris stream in Law et al. (2005; ``L05" hereafter).  
The orbit was constrained to match the current position, radial velocity (RV) and $\mu$ 
(Gardiner \& Noguchi 1996) of the Small Magellanic Cloud (SMC), with orbital pole $(l,b)=(196,-5)^{\circ}$.\footnote{We do not model the possibly complex interaction between 
the Small and Large Magellanic Clouds 
since we are interested in testing a hypothetical stream with desirable properties.}
All other model parameters were similar to those
in L05.  Each simulation was ``observed" outside a $|b|>15^{\circ}$
ZA, with $\sim10^3$, $10^4$ and $10^5$ tracer stars (apportioned with $\sim90\%$ 
of these in the satellite core and $\sim10\%$ in the tidal tails), 
and with 0, 10 and 20\% random Gaussian distance errors imposed.  The simplest (though not best!) 
analysis of these data is simply to fit a plane and measure its intersection with the 
$X_{GC}$ axis.  With this 0th-order method, even for large samples of stars in dynamically cold
streams (i.e., not that from a 10$^{10}$ M$_{\sun}$ progenitor) and no distance errors, 
relatively large ($<7\%$) systematic errors in $R_{0}$ can remain (Fig.\ 1) because plane-fitting
does not account for the residual precessional twisting of the debris arms.  The direction of
precession (determined by the direction of the stream angular momentum vector) 
drives the sense of the imposed systemic $R_{0}$ error (i.e. closer or farther), 
and random distance errors add additional uncertainties depending 
on details of the stream orientation relative to the ZA. 
A better, now proven method (e.g. L05) is to use N-body modeling to reconstruct 
a given stellar stream; such
modeling can precisely account not only for precession but also for stream dispersion and other 
higher order uncertainties,
which would permit a more accurate identification of the center of the MW potential
for an appropriately oriented stream.




Recent measurements of stellar motions around Sgr A$^*$ have led to dynamical parallaxes 
good to 5\% ($7.94\pm0.42$ kpc; Eisenhauer et al. 2003),
a measurement sure to improve with longer
Sgr A$^*$ field monitoring campaigns.  
Few percent quality trigonometric parallaxes of stars
near the GC will be measured as part of a SIM Key Project.  
In either method, the target stars are reasonably expected to lie at the assumed 
center of the MW potential.
The proposed use of tidal streams to measure $R_{0}$ will provide an interesting test
of this hypothesis, since tidal streams orbit the {\it true} dynamical center of the 
{\it integrated} potential over tens of kiloparsec scales.  A comparison of this
center to the Sgr A$^*$ distance could reveal whether the MW may be a lop-sided spiral
(e.g., Baldwin et al. 1980, Richter \& Sancisi 1994, Rix \& Zaritsky 1995).  
Such lop-sidedness can, in fact, be induced by mergers of large satellites (Walker, Mihos
\& Hernquist 1997).
In principle, three well-measured 
tidal streams can verify whether the GC lies along $(l,b) = (0,0)$, 
since the true GC should lie at a mutual intersection of the three corresponding
stream orbital planes.






\section{Velocity of the Local Standard of Rest}

Despite decades of effort, the local MW rotation rate 
remains poorly known, with measurements 
varying by 25\%.  Hipparcos
$\mu$'s \citep{feast97} suggest that the 
Local Standard of Rest (LSR) velocity
is $\Theta_{\rm LSR}
=(217.5\pm 7.0) (R_0/8)$ km s$^{-1}$ --- i.e. near the IAU adopted value of 220 km s$^{-1}$.
But a more recent measurement of $\mu$ for Sgr A$^*$  
(Reid \& Brunthaler 2004) yields a higher $(235.6\pm1.2) (R_0/8)$ km s$^{-1}$, 
whereas direct HST measurements of the $\mu$'s of bulge stars against
background galaxies in the same field
yield $(202.4\pm20.8) (R_0/8)$ km s$^{-1}$ (Kalirai et al. 2004) and 
$(220.8 \pm 13.6) (R_0/8)$ km s$^{-1}$ (Bedin et al. 2003). 
Of course, these measures (as well as any of those depending on the Oort constants) 
rely on an accurate measure of $R_0$ (\S 1).   The 
solar peculiar motion must also be known, but is a smaller correction 
(e.g., $5.3\pm0.6$ km s$^{-1}$; Dehnen \& Binney 1998).
On the other hand, considerations of non-axisymmetry of the disk yield corrections to
the measurements that suggest  
 $\Theta_{\rm LSR}$ may be as low as $184 \pm 8$ km s$^{-1}$ \citep{om98} or lower
(Kuijken \& Tremaine 1994).
Independent methods to ascertain $\Theta_{\rm LSR}$ are of great value because
it is fundamental to establishing the MW mass scale.

Eventually, as part of a Key Project of SIM, $\Theta_{\rm LSR}$ will be measured directly 
by the absolute $\mu$ of stars near the GC.  Here
we describe an independent method for ascertaining $\Theta_{\rm LSR}$ using halo tidal 
streams that overcomes several difficulties
with working in the highly dust-obscured, crowded GC, 
and one 
also insensitive to $R_0$ (for all reasonable values of the latter).  
The ideal tidal stream for this method is one with an orbital pole lying near the 
$Y_{GC}$ axis.  
The Sgr tidal stellar stream 
not only fulfills this requirement, but its stars, particularly its trailing arm M giants, are 
ideally placed for uncrowded field astrometry at high
MW latitudes, and at relatively bright magnitudes for, and requiring only the most
modest precisions from, SIM.  Indeed, as we show, this method is even within
the grasp of future high quality, ground-based astrometric studies.


It is remarkable that the Sun presently lies within a kiloparsec
of the Sgr debris plane (MSWO).  The pole of the plane, 
$(l_p,b_p)=(272, -12)^{\circ}$, means that the line of nodes of its intersection with the 
MW plane is almost coincident with the $X_{GC}$ axis.  Thus (Fig.\ 2) 
the motions of Sgr stars {\it within} this plane are almost entirely contained in their 
Galactic $U$ and $W$
velocity components, whereas the $V$ motions of stars in the Sgr tidal tails almost entirely 
reflect {\it solar motion}.  To the degree that its $V$
distribution is not completely flat in Figure 2 is due to the slight amount of streaming
motion projected onto the $V$ motions from the 2$^{\circ}$ Sgr orbital plane tilt from $X_{GC}$, compounded by (1) Keplerian variations in the space velocity of stars as a function of orbital 
phase, as well as (2) 
precessional effects that lead to $\Lambda_{\sun}$-variable departures of
Sgr debris from the nominal best fit plane to all of the debris. 
The latter is negligible for trailing debris but is much larger for the leading debris, 
which is on average closer to the GC and dynamically older compared to the trailing debris when 
viewed near the Galactic poles (J05).
In addition, because the leading debris gets arbitrarily close to the Sun (L05), projection
effects make it more complicated to use for the present purposes.  Additional problems
with the leading arm debris, which suggest that more complicated effects have perturbed it 
are discussed in L05 and J05.   In contrast, the
Sgr trailing tail is beautifully positioned fairly equidistantly from us
for a substantial fraction of its stretch across the Southern MW hemisphere (MSWO).
This band of stars arcing almost directly ``beneath'' us within the $X_{GC}$-$Z_{GC}$ plane
provides a remarkable zero-point reference against which to make direct 
measurement of the solar motion
{\it almost completely independent of the Sun's distance from the GC. }
The extensive mapping (MSWO) of the Sgr tails with
Two Micron All-Sky Survey  (2MASS) M giants provides an ideal source list for individual
stellar targets from this $>360^{\circ}$-wrapped, MW polar ring.




L05 used M giant spatial (MSWO) and RV 
data (Majewski et al. 2004) to constrain models of
Sgr disruption, best fitting when a 3.5 $\times 10^8$ M$_{\sun}$ Sgr  
of 328 km s$^{-1}$ space velocity orbits with period 0.85 Gyr and
apo:peri-Galactica of 57:13 kpc.  These models fit what appears to be 
$\sim$2.5 orbits (2.0 Gyr) of Sgr mass loss in M giants. 
The adopted MW potential is smooth, static and given by the sum of a 
disk, 
spheroid and 
 halo described by the axisymmetric function
        $\Phi_{\rm halo}=v_{\rm halo}^2 \ln (R^{2}+[z/q]^2+d^{2})$
where $q$ is the halo flattening, $R$ and $z$ are cylindrical coordinates and $d$ is 
a softening parameter. 
Additional model details
are given in L05.

Model fits to the Sgr spatial and velocity 
data allow predictions of the 6-D phase space
configuration of Sgr debris.  Figure 2 shows predicted $U,V,W$ velocity
components of debris as a function of longitude, $\Lambda_{\sun}$, in the Sgr orbital 
plane (see MSWO).
The debris is shown assuming 
$q=0.9$, $R_{0} = 7$ kpc, and a characterization
of the total potential whereby the LSR speed is
$\Theta_{\rm LSR}$ = 220 km s$^{-1}$.  L05 explores how variations in $q$
affect primarily the $U$ and $W$ (through projection along
RV).  
Figure 3 (green, yellow, and magenta points) shows how variations in the {\it scale} of the 
potential, expressed through variations in adopted $\Theta_{\rm LSR}$, 
affect $V$.  Clearly, $\Theta_{\rm LSR}$ ranging from
180 to 260 km s$^{-1}$ translates to obvious variations in observed $V$ for
trailing arm stars.  This effect is easily separable from any residual uncertainty in the shape of the
potential or $R_{0}$:
Figure 3 (red and blue points respectively) 
illustrates negligible $V$ changes produced by holding $\Theta_{\rm LSR}$ fixed
at 220 km s$^{-1}$ but varying 
$q$ from 0.9 to 1.25 (i.e. oblate to prolate) and $R_0$ from 7 to 9 kpc.


Figure 3 is the basis for the proposed use of Sgr to measure $\Theta_{\rm LSR}$.  Ideally,
to execute the experiment requires obtaining $V$ from the observed
$\mu$ and RVs of Sgr arm stars.  However, 
because of the particular configuration of Sgr trailing arm debris, almost all of $V$ is
reflected in the $\mu$ of these stars, and, more specifically,  the reflex
solar motion is contained almost entirely
in the $\mu_l \cos(b)$ component of $\mu$ for Sgr trailing arm stars
away from the MW pole.  Working in the observational, 
$\mu$ regime means
that vagaries in the derivation of {\it individual} star distances can be removed from the
problem, as long as the system is modeled with a proper {\it mean} distance for the
Sgr stream as a function of $\Lambda_{\sun}$.
Figure 4
shows three general regimes of the trailing arm $\mu_l \cos(b)$ trend:
(1) $\Lambda_{\sun}  \gtrsim100^{\circ}$ where $\mu_l \cos(b)$ is positive and roughly
constant, (2) the region from $100^{\circ} \gtrsim \Lambda_{\sun} \gtrsim 60^{\circ}$
 where $\mu_l \cos(b)$ flips sign as the debris
passes through the South Galactic Pole to shift the Galactic longitudes of the trailing arm
by $\sim180^{\circ}$, and (3) 
$\Lambda_{\sun} \lesssim 60^{\circ}$, where $\mu_l \cos(b)$is negative 
and becomes smaller with decreasing $\Lambda_{\sun}$ (because the Sgr stream becomes
increasingly farther).  The sign flip in $\mu_l \cos(b)$ is a useful happenstance
in the case where
one has $\mu$ data not tied to an absolute reference frame but which is at 
least robust to systematic zonal errors: In this case the 
peak to peak amplitude of $\mu_l \cos(b)$ for the trailing arm stars yields (two times) the 
reflex motion of the Sun\footnote{We find that these peaks lie at $\Lambda_{\odot} = 60-65^{\circ}$
and $115-120^{\circ}$; note that technically $V \propto (\mu/d)_{60-65} + (\mu/d)_{115-120}$.}.


The intrinsic RV dispersion of the Sgr trailing arm has been measured to be
$\sim$10 km s$^{-1}$ (Majewski et al. 2004); assuming symmetry in the two 
transverse dimensions of the stream gives an intrinsic $\mu$ dispersion of the Sgr
trailing arm of $\sim$0.1 mas yr$^{-1}$ (see Fig.\ 4a).  Thus, until SIM-quality proper 
motions exist, the measurement of the reflex solar motion by this method will be dominated by 
the error in $\mu$.  To quantify the accuracy of the proposed method, we 
introduce artificial random errors into the proper
motions of the five models shown in Figure 3 and calculate the accuracy with which we
expect to recover the solar reflex motion.

Simply applying the formalism described above, we recover $\Theta_{\rm LSR}$ values 
of\footnote{Correcting
for the assumed 12 km s$^{-1}$ speed of the Sun with respect to the LSR.}
212, 255, and 279 km s$^{-1}$ in models for which input 
$\Theta_{\rm LSR} = 180, 220,$ and 260 km s$^{-1}$ respectively.  This indicates
that the method systematically overpredicts $\Theta_{\rm LSR}$  
by about 30 km s$^{-1}$; this is because (see Fig.\ 3) the trend of $V$ with $\Lambda_{\odot}$ 
is not perfectly flat but changes by $\sim30$ km s$^{-1}$ between the
peaks at $\Lambda_{\odot} =60-65^{\circ}$ and 115-120$^{\circ}$.  Correcting for this systematic
bias, we perform 1000 tests where we 
randomly draw particles from the model debris streams in these ranges with artificially added random
scatter in the $\mu$, and find that recovering the solar velocity to within 10 km s$^{-1}$ requires a sample
of approximately 200 stars with $\mu$ measured to about 1 mas yr$^{-1}$ precision with no zonal systematics.  Using the models with $\Theta_{\rm LSR} = $ 180/220/260
km s$^{-1}$, these tests recover mean values of 182/225/249 km s$^{-1}$ respectively 
with a dispersion of results between the tests of 10 km s$^{-1}$.
As expected from Figure 3, varying $q$ and $R_{0}$ has negligible effect: 
Tests on models in both of these MW 
potentials (where $\Theta_{\rm LSR} = 220$ km s$^{-1}$) recover mean values of 225 and 228 km s$^{-1}$.


Present astrometric catalogs are just short of being able to do this experiment:
Hipparcos is not deep enough, the Southern Proper Motion Survey 
(Girard et al. 2004) has not yet covered enough appropriate sky area,
and UCAC2
(Zacharias et al. 2004) has several 
times larger random errors than useful as well as comparably-sized
zonal systematic errors at relevant magnitudes 
(N. Zacharias, private communication).  However, 
to demonstrate how only modest advances
in all-sky $\mu$ precisions are needed to make a definitive measurement, 
Figure 4 includes a direct comparison of the $\mu_l \cos(b)$ trend for Sgr 
M giants using UCAC2 $\mu$'s for 2MASS M giants.  Impressively, the overall expected
$\mu_l \cos(b)$ trends can be seen, but the large scatter and systematic shifts in the trailing arm
motions belie the limits of UCAC2 accuracies at $V \sim 15$.  Even a factor of two 
improvement in UCAC2 random errors {\it and} elimination of zonal errors might
lead to a useful measurement of $\Theta_{\rm LSR}$.  It is not unreasonable to expect
advances in all-sky $\mu$ catalogues at this level soon (e.g., from the
Origins Billion Star Survey or Gaia), but in any case SIM PlanetQuest
will {\it easily} obtain the necessary $\mu$ (and parallaxes) of selected 
Sgr trailing arm giants.


\acknowledgements 
We appreciate funding by NASA/JPL through the Taking Measure of the MW
Key Project for SIM PlanetQuest, NSF grant AST-0307851,
the Packard Foundation, and the F.H. Levinson Fund of the
Peninsular Community Foundation.  SRM appreciates the hospitality of the Carnegie Observatories
during the writing of this paper.

\newpage

\begin{figure}[!ht]
\includegraphics[angle=90]{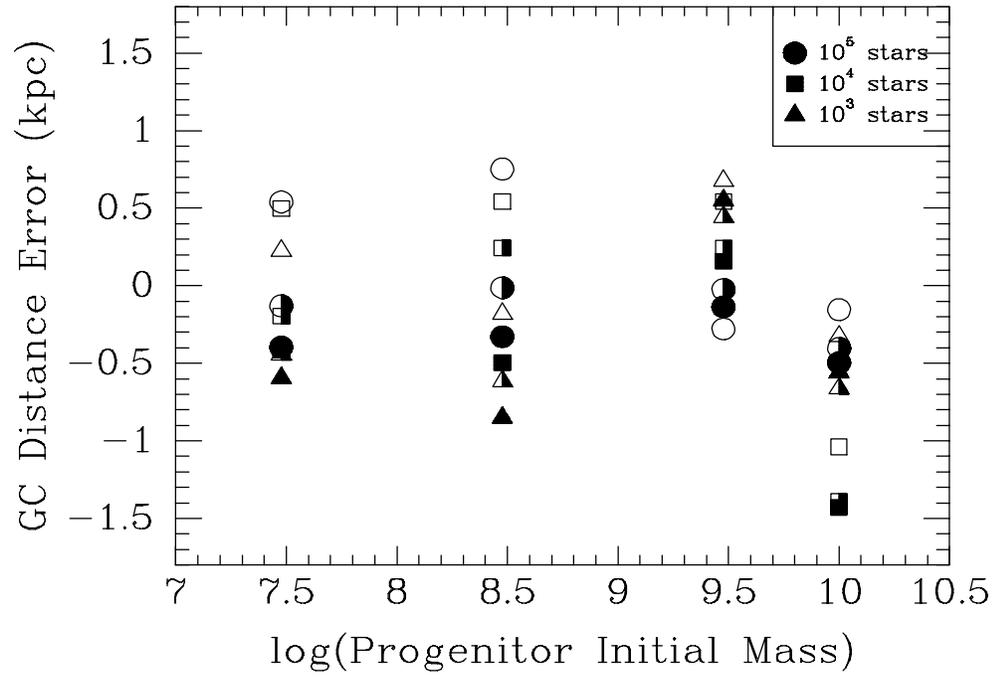}
\caption{Results of plane-fitting to simulations of disrupting satellites of different masses
orbiting similarly to the Magellanic Clouds.  Filled,
half-filled and open symbols are for simulations with no, 10\% and 20\% distance errors 
imposed on the member stars.  }
\end{figure}

\begin{figure}[!ht]
\plotone{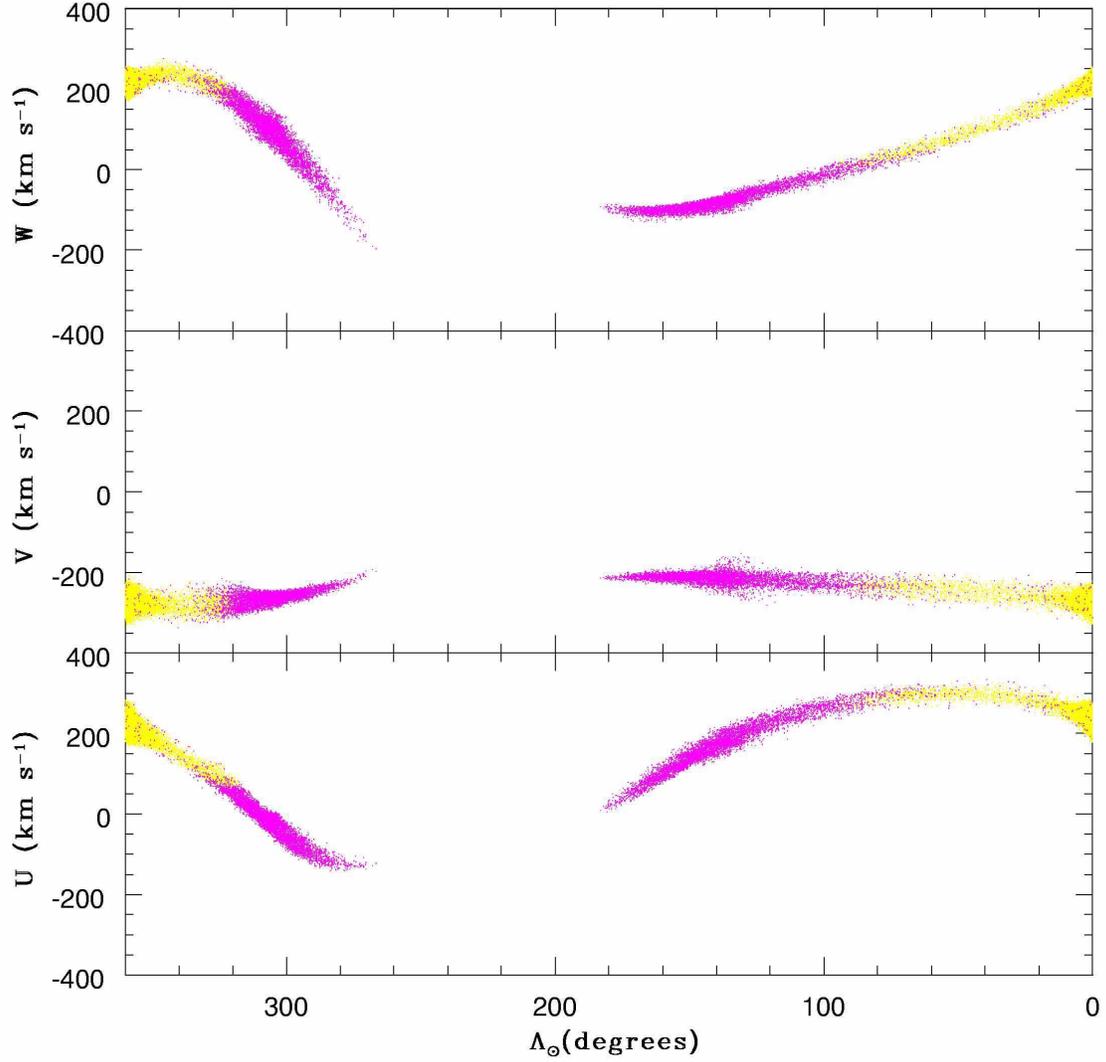}
\caption{Predicted $U,V,W$ velocities (right-handed system) as a function of longitude in the Sgr orbital plane
($\Lambda_{\sun}=0^{\circ}$ at present Sgr position).
Debris lost on last half (yellow) and previous full (magenta) orbits are shown (see Law et al. 2005).
Trailing debris
stretches from $\Lambda_{\sun}=0^{\circ}$ to $\sim
180^{\circ}$; leading arm debris goes from $\Lambda_{\sun}=0^{\circ}$ to $ \sim 270^{\circ}$.  The Galactic model has a $q=0.9$ halo with $\Theta_{\rm LSR} = $ 220 km s$^{-1}$ and $R_0 =$ 7 kpc.}
\end{figure}

\begin{figure}[!ht]
\plotone{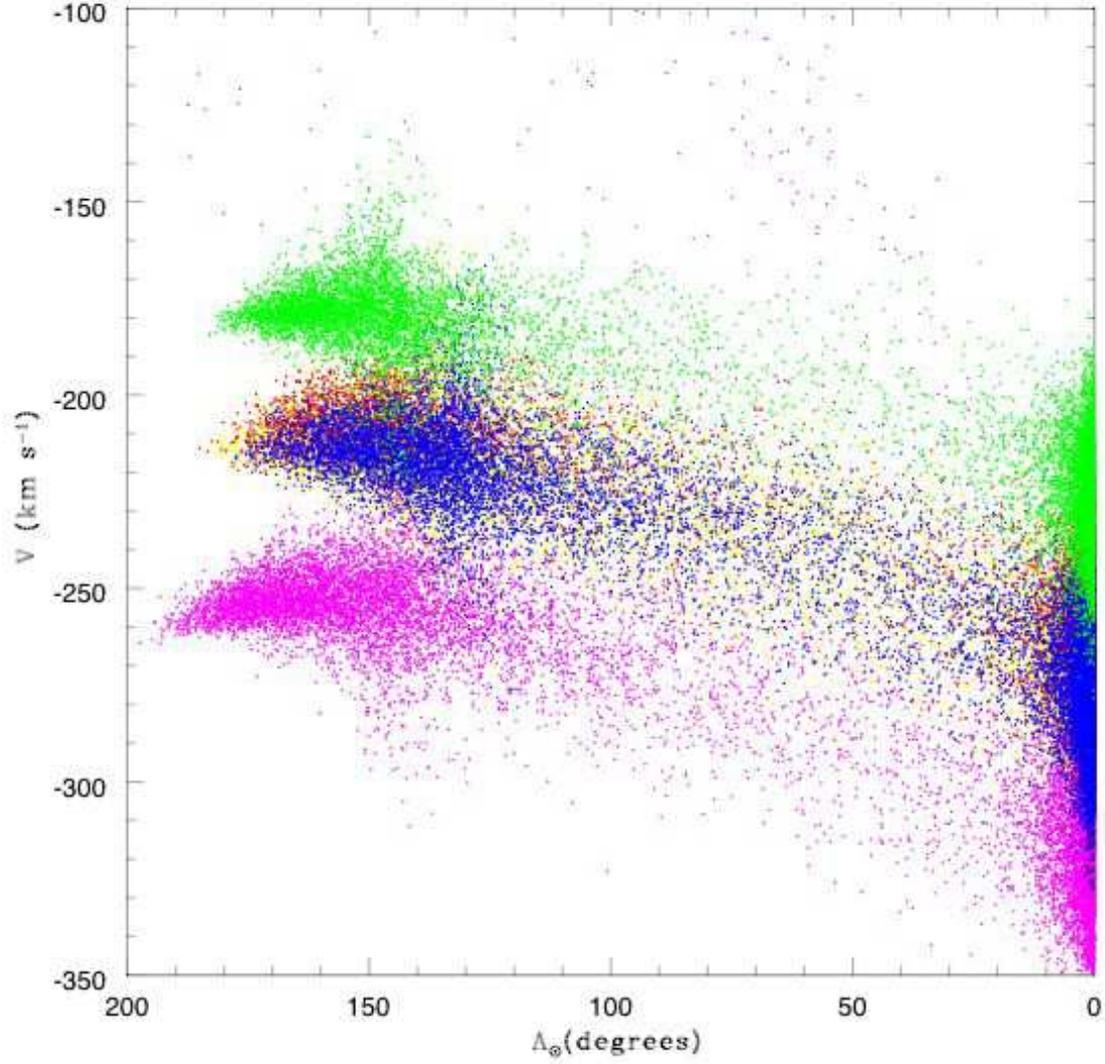}
\caption{Variation of the Sgr trailing arm $V$ velocities for a range of $\Theta_{\rm LSR}$,
 $R_0$, and $q$.
Green, yellow, and magenta points represent debris from satellites disrupting in potentials where $q = 0.9$ and $R_0 =$ 7 kpc, but $\Theta_{\rm LSR}$ is 
180, 220, and 260 km s$^{-1}$, respectively.  Red and blue points represent satellite debris
in potentials where $\Theta_{\rm LSR} =$ 220 km s$^{-1}$ but $q = 1.25$ ($R_0$ fixed at 7 kpc) and $R_0 = 9$ kpc
($q$ fixed at 0.9), respectively.  Note how changes in the \textit{shape} of the potential have little to no
effect on $V$, while changes in the \textit{scale} of the potential produce large, 
approximately linear shifts.}
\end{figure}

\begin{figure}[!ht]
\plotone{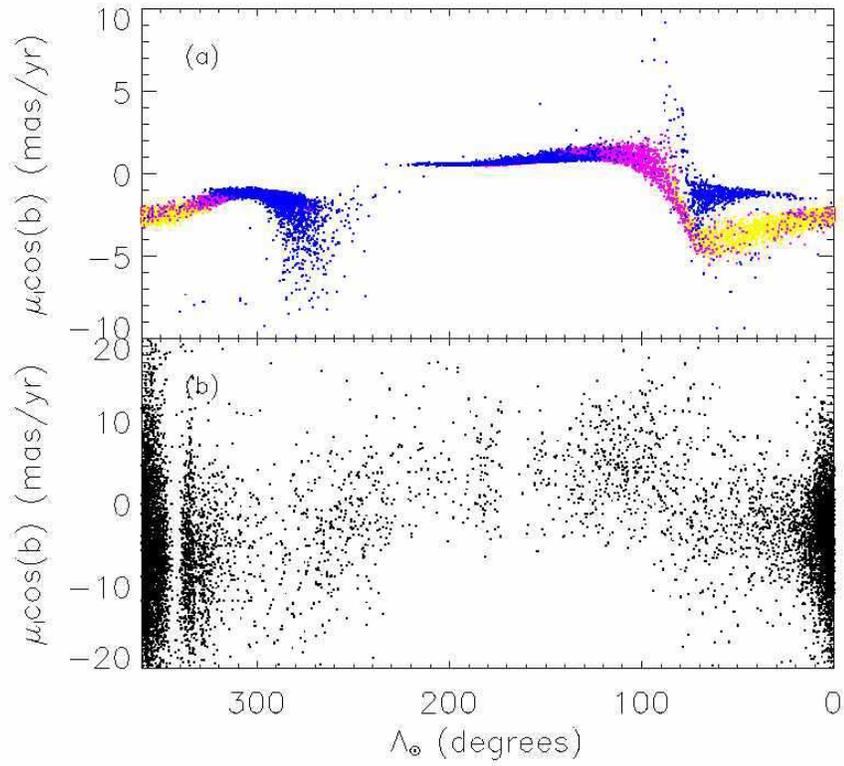}
\caption{(a) Same as Fig.\ 2 (i.e.,same MW model), but for 
predicted $\mu_l \cos(b)$ and for 2.5 orbits of mass loss.
(b) Observed UCAC2 $\mu_l \cos(b)$ for 2MASS M giants in Galactic regions dominated
by Sgr stream stars (all M giants within 7 kpc of the nominal Sgr plane and having
heliocentric distances of 15-30 kpc).  Note the differing vertical scales of the two panels.
}

\end{figure}

\end{document}